\documentclass[twocolumn,showpacs,preprintnumbers,amsmath,amssymb]{revtex4}

\usepackage{graphicx}
\usepackage{dcolumn}
\usepackage{bm}

\begin{document}
\title{Ground Band and a Generalized GP-equation for Spinor Bose-Einstein
Condensates}
\author{C.G. Bao$^{1,2}$ and Z.B. Li$^{1}$}

\affiliation{$^{1}$State Key Laboratory of Optoelectronic
Materials and Technologies,} \affiliation{\ and Department of
Physics, Zhongshan University, Guangzhou 510275, P.R. China, }
\affiliation{\ $^{2}$Center of Theoretical Nuclear Physics,
National Laboratory of Heavy Ion Accelerator, Lanzhou 780000, P.R.
China }

\vspace{1pt}
\begin{abstract}
\qquad For the spinor Bose-Einstein condensates both the total spin $S$ and
its Z-component $S_{Z}$\ should be  conserved. \ However, in existing
theories, only the conservation of $S_{z}$\ has been taken into account. \
To remedy, this paper is the first attempt to take the conservation of both $%
S$ and $S_{Z}$ into account. \ For this purpose, a total
spin-state with the good quantum numbers  $S$ and $S_{Z}$ is
introduced in the trial wave function, \ thereby a generalized
Gross-Pitaevskii equation has been
derived. \ With this new equation, the ground bands of the  $^{23}$Na and $%
^{87}$Rb condensates have been studied, where the levels distinct in $S$
split. \ It was found that the level density is extremely dense in the
bottom of the ground band of $^{23}$Na, \ i.e., in the vicinity of the
ground state. \ \ On the contrary, for $^{87}$Rb, the levels are extremely
dense in the top of the ground band,

\end{abstract}

\pacs{03.75.\ Fi, \ 03.65.\ Fd}
\maketitle

\vspace{1pt}

1, \textbf{Introduction}

In 1998 the spinor Bose-Einstein condensation of sodium atoms was
experimentally realized in an optical trap by Stamper-Kurn
\textit{et al}\ \cite{stam98}. The advantage of the optical trap
is that it liberates the spin degrees of freedom of atoms (e.g., \
$^{23}$Na, $^{39}$K, and $^{87}$Rb atoms have nuclear spin 3/2 and
electrons at $s$ orbits, they behave as simple bosons with spin
one at low temperature). As a consequence of the liberation, such
a macroscopically coherent system with internal degrees of freedom
becomes very rich in physics. Several interesting phenomena have
already been obtained, e.g., quantum entanglement of spins\ \cite
{cira98,duan02}, spinor four-wave mixing\ \cite{law98, gold98},
and super fragmentation and coherent fragmentation\ \cite{ho00}.

For the spinor Bose-Einstein condensates of bosons with spin one,
the pairwise interaction $U_{ij}$ is spin-dependent. It can be
written as\ \cite{ho98, ohmi98, zhan98, kemp02, crub99}
\begin{eqnarray}
U_{ij}&=&\delta (\mathbf{r}_{i}\mathbf{-r}_{j})(g_{m}P_{m}+g_{q}P_{q})  \nonumber \\
&=&\delta (%
\mathbf{r}_{i}\mathbf{-r}_{j})(c_{0}+c_{2}\mathbf{F}_{i}\cdot \mathbf{F}_{j})
\label{Uij}
\end{eqnarray}
where $P_{m}$ and $P_{q}$ \ are the projectors to two-body spin-states with
two-body spin equal to 0 and 2, respectively, and $g_{m}$ and $g_{q}$\ are
the associated strengths. $\mathbf{F}_{i}$ is the spin operator of particle $%
i$, $c_{0}=(g_{m}+2g_{q})/3$\ and $c_{2}=(g_{q}-g_{m})/3$. \ The term with $%
c_{0}$ is a central interaction, while the term with $c_{2}$ is
spin-dependent.

When $g_{q}\approx g_{m}$\ (e.g., for $^{87}$Rb), we have $c_{2}<<c_{0}$ and
therefore the spin-dependent term is unimportant. In this case the\
single-mode approximation (SMA) works very well\ \cite{duan02, law98,
gold98, ho00, koas00}. In SMA, a common mode function $\phi (\mathbf{r})$ is
obtained by solving the Gross-Pitaevskii (GP) equation with the central
interaction alone, while the small spin-dependent interaction is treated as
a perturbation.

When $c_{2}$ is not very small, the perturbation treatment is
poor. In this
case, a three-mode approximation (TMA) is proposed. Instead of $\phi (%
\mathbf{r})$, \ three mode functions $\phi _{\mu }(\mathbf{r})$ are
introduced, where $\mu =1,0,$ and $-1$ are associated with the three
components of spin. These functions obey a set of coupled GP equations with
all the interactions taking into account\ \cite{yi02}.

In the above theories the total spin $S$ of the condensate is not
explicitly considered but only its Z-component $S_{Z}$. \ However,
the interaction (eq.(\ref{Uij})) assures that $S$ is exactly a
good quantum number. In fact, the eigenstates should be
$S$-dependent but $S_{Z}$-independent (if external magnetic field
is not applied) . Therefore, a more perfect theory should preserve
$S$ . \ It is the aim of this paper to introduce a $S$-conserved
description for spinor condensates. \ For this purpose, we need
the knowledge from few-boson systems. \ This is discussed in the
next section.

\vspace{1pt}

2, \textbf{Few-boson model systems}

Let us first inspect some few-boson 2-dimensional model systems with spin
one and mass $m$, and they are confined by a harmonic trap with frequency $%
\omega $. When $\hbar \omega $ and $\sqrt{\frac{\hbar }{m\omega }}$ are used
as units of energy and length, respectively, the Hamiltonian of the $N$%
-boson system reads
\begin{equation}
H=\frac{1}{2}\sum_{i}(-\nabla _{i}^{2}+r_{i}^{2})+\sum_{i<j}U_{ij}
\label{Ham}
\end{equation}
\ Using the method as given in the ref.\cite{bao03,bao02} , the Hamiltonian
for $N=2$, $3$, and $4$ \ were diagonalized. \ Of course, both the orbital
angular momentum $L$\ and $S$ are good quantum numbers. \ However only the $%
L=0$ states are considered here. \ For a series of eigenstates with a given $%
S,$ let the lowest one be called a first-state. \ Let the energy and wave
function of a first-state be denoted as $E_{S}$ and \ $\Psi _{S}$\ . \ When $%
g_{q}+g_{m}=0.4$\ is assumed, the evolution of $E_{S}$ \ in accord with $%
\beta =g_{q}/g_{m}$ is given in Fig.\ 1. In this figure, there are three
groups of curves for the three cases $N=2,3,$ and $4,$ respectively.\ \ In
each group the curves have distinct $S$\ but the same $S_{Z}=0$. \ When $%
\beta =1$\ , the interaction is spin-independent and the ground
state is degenerate against $S$. \ When $\beta $ deviates from 1,
the interaction depends on the spin and, accordingly, the ground
state splits into a band, the ground band. \ For an example, \
when \ $N=4$, the ground band contains three members with $S=0,2$,
and $4.$ \ When \ $\beta <1$, a pair of boson would be less
repulsive if their spins are coupled to 2 instead of 0 (because \
$g_{q}<g_{m}$). \ Since in $\Psi _{4}$ any pair of bosons must
couple to 2, this state is lower than $\Psi _{2}$ and $\Psi _{0}$.
\ On the contrary, when \ $\beta >1$, \ $\Psi _{4}$ is the highest
state of the band. \ Since the split of the ground state arises
from the character of the interaction, it would remain when $N$\
is large. \ In this case the spins will have many ways to couple
with each other to form total spin-states with $S$ ranging from 0
(or 1) to $N.$ \ It implies that , when $N$\ \ is large, the
ground band would contain many members distinct in their total
spin-states. \ To clarify the structure of the ground band is
another aim of this paper. \ Incidentally, since the members of
the band are already orthogonal due to their spin structures,
their spatial wave functions may be alike.

Let us analyze the wave functions to see the structure of the
ground band of the few-boson systems. In the $S$-conserved form,
the wave function is written as
\begin{equation}
\Psi _{S}=\sum_{\lambda i}f_{S\lambda i}\mathbf{\theta
}_{S,\lambda i} \label{PsiS}
\end{equation}
where $\theta _{S,\lambda i}$ is the total spin-state with the good quantum
number $S$, and it is also the $i-th$ basis function of the $\lambda $
representation of the permutation group (e.g., refer to ref.\lbrack
15\rbrack ). \ $f_{S\lambda i}$ is a function of spatial coordinates
belonging to the same representation. The summation over $i$\ is necessary
to assure $\Psi _{S}$ to be totally symmetric. \ The summation over $\lambda
$\ responds to the possibility that some total spin-states distinct in $%
\lambda $\ may have the same $S$ (e.g., for $N=3$ and $S=1,$ there are three
total spin-states \ $\mathbf{\theta }_{1,\{3\}1},$ $\mathbf{\theta }%
_{1,\{21\}1},$ and $\mathbf{\theta }_{1,\{21\}2}$\ )$^{15}$. \
From eq.(\ref{PsiS}), for a $\Psi _{S}$, we can define the weight
of a representation as
\begin{equation} W_{\lambda}=\sum_{i}<f_{S\lambda i}|f_{S\lambda i}>
\end{equation}

It turns out, from our numerical results, that all the states of
the ground band are dominated by $\lambda =\{N\}$\ component. \
E.g., when $N=4,$ the second lowest state is $\Psi _{2}$\ (cf.
Fig.1). \ For $S=2$, there are six total spin-states belonging to
the \{4\}, \{31\}, and \{22\} representations, respectively. \
When $\beta =0.6$ , and $g_{m}+g_{q}$ varies in the range from 0
to 3, our calculation shows that the $W_{\{4\}}$\ of $\Psi _{2}$
ranges from 1 to 0.996, \ it implies that the symmetry \{4\} is
dominant. \ The increase of $g_{m}+g_{q}$ causes only a very small
decrease of $W_{\{4\}}$ from the unity. \ The symmetries other
than \{4\} are found to be important only in excited states but
not in the ground band. \ This holds also when $\beta $\ is given
at other values. \ The underlying reason is that the $\lambda $\
of the spatial wave function must be the same as the total
spin-state. \ If $\lambda \neq \{N\},$ all the bosons can not
condensate to the same spatial state, \ at least a boson should be
spatially excited. \ This excitation will cause an explicit
increase of energy. \ Since the ground band is an aggregation of
the first-states, each of them will do their best to lower the
energy. \ Therefore the \{N\}-symmetry is pursued. \ Evidently,
the above explanation remains valid not matter how large $N$ is.\
\ Thus, it is expected that the ground band of a large-$N$ system
remains to be dominated by $\lambda =\{N\}$.

\vspace{1pt}

3, \textbf{Symmetrized total spin-states}

Let us now consider the ground band of the systems with many
bosons\ . \ Based on the above discussion, only the case $\lambda
=\{N\}$ is necessary to be considered. As a first step, we have to
clarify the number of symmetrized total spin-states. For a
$N-boson$ system, let $\theta _{S_{Z}}^{,j}$ \ denotes a
normalized totally symmetric spin-state having only\ $S_{Z}$
conserved, \ \ where the superscript $j$ is introduced to number
this kind of states. \ For a given $S_{Z}$, \ let the total number
be denoted as $M_{Z}(S_{Z}),$ we have

\vspace{1pt}
\begin{equation}
M_{Z}(S_{Z})=\frac{1}{2}(N-S_{Z}-\frac{1-(-1)^{N-S_{Z}}}{2})+1  \label{Mz}
\end{equation}

On the other hand, let $\theta _{S}^{k}$ \ denotes those having\
both $S$ and $S_{Z}$ conserved. \ For a given $S$, let the total
number of $\theta _{S}^{k}$ \ be denoted as $M(S),$ which\ is
related to $M_{Z}(S_{Z})$\ as
\begin{equation}
M(S)=M_{Z}(S)-M_{Z}(S+1)=(1+(-1)^{N-S})/2  \label{M}
\end{equation}
From (\ref{M}) we know that \ $M(S)$\ is \textit{one} if $N-S$ is even, or is%
\textit{\ zero }if $N-S$ is odd. Therefore, the superscript $k$ is
redundant, and a simpler notation $\theta _{S}$ \ is used hereafter. \
Evidently, $\theta _{S}$ is a simpler notation of $\theta _{S,\{N\}1}$
defined before. \ \ For the calculation of matrix elements, the following
formula is useful.
\begin{eqnarray}
<\theta _{S}|\mathbf{F}_{1}\mathbf{.F}_{2}|\theta _{S}>&=&\frac{2}{N\;(N-1)}%
<\theta _{S}|\sum_{i<j}\mathbf{F}_{i}\mathbf{.F}_{j}|\theta _{S}> \nonumber \\
&=&\frac{%
S(S+1)-2N}{N\;(N-1)}
\end{eqnarray}
Where the first equality holds because $\theta
_{S}$\ is totally symmetric, the second equality holds because
\begin{equation}
2\sum_{i<j}\mathbf{F}_{i}\mathbf{.F}_{j}=\;(\sum_{i}\mathbf{F}%
_{i})^{2}-\sum_{i}\mathbf{F}_{i}^{2}
\end{equation}

When $\theta _{S}$\ are used for the spin-states, the spatial wave
functions must be totally symmetric just as those for the scalar
bosons. \ Therefore, in each member of the ground band,
condensation would occur just as in the scalar system disregarding
what $S$\ is. \ Thus, as a reasonable approximation, $\Psi _{S}$\
that has been given in eq.(\ref{PsiS}) is now simplified as
\begin{equation}
\Psi _{S}=\Pi _{i=1}^{N}\phi _{S}(\mathbf{r}_{i})\;\theta _{S}  \label{PsiS2}
\end{equation}
\ \ Where the subscript in $\phi _{S}$ responds to the fact that the spatial
states may depend on $S$. Incidentally, the functions $\phi _{S}(%
\mathbf{r})$\ with distinct $S$\ are in general not mutually orthogonal.

\vspace{1pt}

4, \ \textbf{The generalized Gross-Pitaevskii equation}

\vspace{1pt}Let us insert (\ref{PsiS2}) into the three-dimensional
Schr\"{o}dinger equation
\begin{equation}
H\Psi _{S}=E_{S}\Psi _{S}  \label{Schr}
\end{equation}
where $H$\ is given in (\ref{Ham}). \ Using the standard variational
procedure, a generalized \ Gross-Pitaevskii (GP) equation can be derived as
\begin{equation}
\lbrack \frac{1}{2}(-\nabla ^{2}+r^{2})+(N-1)g_{S}\;\rho _{S}\rbrack \phi
_{S}(r)=\varepsilon _{S}\;\phi _{S}(r)  \label{GP}
\end{equation}
where
\begin{eqnarray}
g_{S}&=&c_{0}+<\theta _{S}|\mathbf{F}_{1}\mathbf{.F}_{2}|\theta
_{S}>c_{2} \nonumber \\
&=&c_{0}+\frac{S(S+1)-2N}{N\;(N-1)}c_{2}
\end{eqnarray}

$\phi _{S}(r)$ is normalized and isotropic, $\rho _{S}=\phi
_{S}(r)^{\ast }\phi _{S}(r)$ and $\varepsilon _{S}$ is related to
the eigenenergy $E_{S}$\ as
\begin{equation}
E_{S}=N\;\varepsilon _{S}-\frac{N\;(N-1)}{2}g_{S}\stackrel{\_}{\rho _{S}}
\label{Es}
\end{equation}
where
\begin{equation}
\stackrel{\_}{\rho _{S}}=\int d\mathbf{r}\phi _{S}(r)^{\ast }\rho _{S}\phi
_{S}(r)  \label{rhobar}
\end{equation}

The generalized GP equation is the same as the usual one with only one
exception, namely the strength $g_{S}$ is not a constant but $S-$ and $N-$%
dependent. Consequently, the bosons in each state of the ground band
condensate into a specific single-particle states $\phi _{S}$ depending on $%
S $. \ \ From (\ref{M}) we know that the band contains totally
$N/2$+1 \ states if $N$ is even with
their $S$ ranging from 0, 2, 4, to \ $N$, \ or contains $(N+1)/2$ states if $%
N$ is odd with $S$ ranging from 1, 3, 5, to $N$. \ When $\beta
=1$, the generalized GP equation will automatically reduce to the
usual GP equation.
When $\beta >1$, $c_{2}$ is positive,\ a larger $S$ will lead to a larger $%
g_{S}$ , and therefore a higher level. Thus, the upmost level of the ground
band has $S=N$\ , while the ground state has $S=0$. \ On the contrary, when $%
\beta <1$, \ the ground state has $S=N$. \ This finding coincides with
previous findings\ \cite{law98}.

Let us go to realistic cases (three-dimensional) with a large N to
inspect the structure of
the ground band. For \ $^{23}$Na atoms, we have $g_{m}=6.351\times 10^{-4}%
\sqrt{\omega }$ and $\beta =1.10$ \ . For \ $^{87}$Rb atoms, we have $%
g_{m}=2.52\times 10^{-3}\sqrt{\omega }$ and $\beta =0.986$. \ By solving (%
\ref{GP}), the ground state energies $E_{g}$ per particle with respect to $%
N\omega ^{1/2}$ are plotted in Fig.\ 2\ . \ \ In order to see the
broadening of the ground band, $(E_{S}-E_{S_{min}})/N$ \ with
respect to $S/N$ \ is plotted in Fig.\ 3\ , where $\omega =100Hz$
is given, the subscript $S_{min}=0$ (if $N$ even), or $=1$ (if $N$
odd).

Now, let us evaluate the level-density. \ When $S/N<<1,$ eq.
(\ref{GP}) can be solved perturbatively (the perturbation is
caused by $g_{S}-g_{S_{min}}$), we obtain
\begin{equation}
\varepsilon _{S}-\varepsilon _{S_{min}}=(N-1)(g_{S}-g_{S_{min}})\stackrel{\_%
}{\rho}_{S_{min}}=c_{2}\;\frac{S(S+1)}{N\;}\stackrel{\_}{\rho
}_{S_{min}}
\end{equation}

Thereby the difference $E_{S}-E_{S_{min}}$\ can be obtained, and
we have the level-density
\begin{eqnarray}
D(E_{S})&=&|\frac{dn}{dE_{S}}|=1/(|c_{2}|\stackrel{\_}{\rho }_{S_{min}}%
(2S+1)) \nonumber \\
&=&S(S+1)/(2|E_{S}-E_{S_{min}}|(2S+1))
\end{eqnarray}
where $dn$ is the number of levels in $dE_{S}$. \ Since $%
|E_{S}-E_{S_{min}}|$ is a very small quantity in the vicinity of
$E_{S_{min}}$, \ the
levels are extremely dense in the vicinity of the ground state (if $\beta >1$%
) or in the top of the ground band (if $\beta <1$). \ This is an interesting
point.

When $S/N$\ is not very small, via a numerical evaluation with a high
precision, $E_{S}$ can be represented by a parabolic curve
\begin{equation}
E_{S}=E_{S_{min}}\pm N(\frac{S}{N})^{2}d  \label{Esappr}
\end{equation}
where $d$ is a function of $N$ \ as given in the inset of Fig.\
3\ , the positive (negative) sign is for $\beta >1\;(<1)$. \ The
band width $E_{bw}=$ $|E_{N}-E_{S_{min}}|$ \ is given in Fig.\ 4\
versus $N\omega ^{1/2}$ . \ The level-density is then
\begin{equation}
D(E_{S})=\frac{1}{4}\sqrt{\frac{N}{|E_{S}-E_{S_{min}}|\;d}}
\label{dol}
\end{equation}
They are given in Fig.\ 5,\ where $\omega =$ 100Hz is assumed.

If the generalized GP equation is not solved exactly but perturbatively (the
perturbation is caused by $g_{S}-c_{0}$ ), then the resultant energies are
identical to those given by the SMA$^{4}.$

\vspace{1pt}

5,\textbf{\ Final remarks}

In conclusion, we have derived a generalized GP equation based on
a $S$ -conserved formalism . This equation can describe not only
the ground state but the whole ground band. \ In the band the
states having larger $S$ are higher (lower) in $E_{S}$ \ if $\beta
>(<)\;1$ . \ It is recalled that $S$ is in fact a good quantum
number. The present theoretical framework satisfies this
requirement, and has a very simple formalism to facilitate
numerical calculations. Furthermore, (\ref{PsiS2}) is the only
approximation, which is comparable to that used for deriving the
well known GP equation. Therefore, the reliabilities of these two
equations are comparable. Thus, this work is an improvement of the
SMA and TMA, while at the same time avoids perturbative treatment
or the difficulty of solving coupled equations.

Acknowledgment: We appreciate the support by NSFC under the grants 90103028
and 90306016\ .

\clearpage

\begin{figure}
\includegraphics{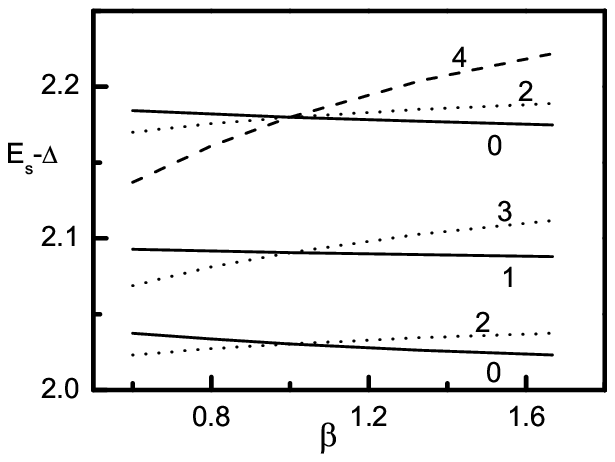}
\caption{\label{fig:1} The evolution of $E_{S}$\ \ (in $\hbar
\omega $\ ) with $\beta $ in few -boson systems. \ Where the
lowest two curves, the two in the middle, and the upper three are
for 2-, 3- and 4-boson systems, respectively. \ $S$ is marked by
the Curves. $\Delta$ is introduced to shift the curves,
$\Delta=0,1$, and $2$ for $N=2,3$, and $4$ systems, respectively.}
\end{figure}

\begin{figure}
\includegraphics{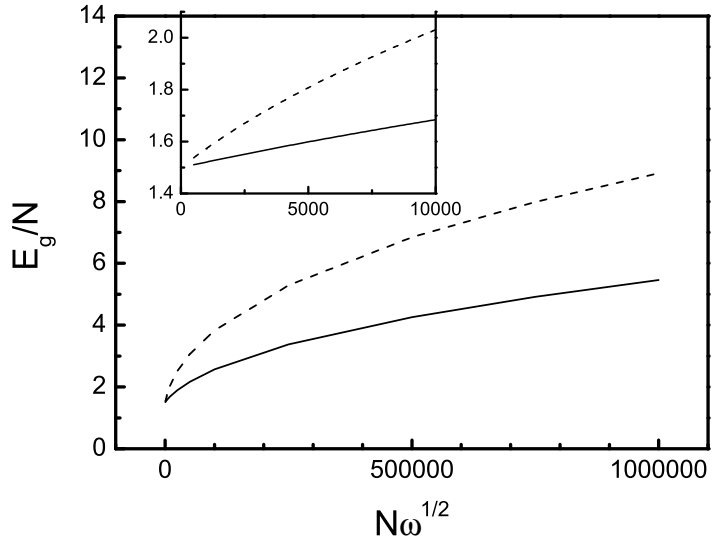}
\caption{\label{fig:2} The ground state energy $E_{g}$ per
particle (in $\hbar \omega $) versus $N\omega ^{1/2} $. \ The
total spin of the ground state is 0 or 1 (if $^{23}$Na), or is $N$
(if $^{87}$Rb). In Fig.\ 2 to 5 the solid (dashed) lines are for
$^{23}$Na ($^{87}$Rb). \ The inset is just for smaller $N\omega
^{1/2}$\ .}
\end{figure}

\begin{figure}
\includegraphics{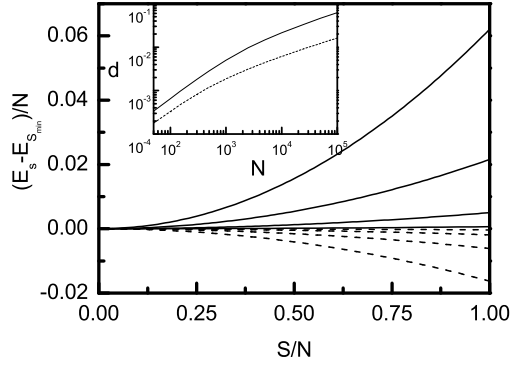}
\caption{\label{fig:3} $(E_{S}-E_{S_{min}})/N$ (in $\hbar \omega
$)\ versus $S/N$ with $\omega =100Hz$\ . The four
solid (dashed) lines from the lowest to highest (highest to lowest) are for $%
N=10^{2},\;10^{3},\;10^{4}$, and\;$10^{5}$, respectively. The
inset is $d$ versus $N$.}
\end{figure}

\begin{figure}
\includegraphics{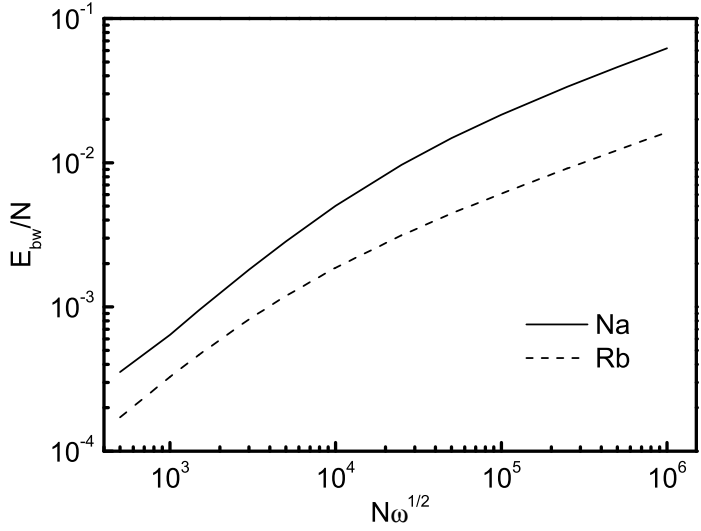}
\caption{\label{fig:4} The band width $E_{bw}/N$ (in $\hbar \omega
$)\ versus $N\omega ^{1/2}$\ .}
\end{figure}

\begin{figure}
\includegraphics{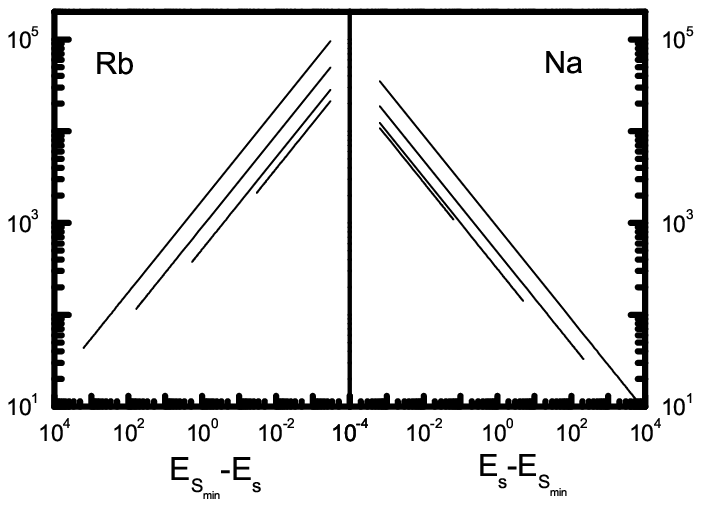}
\caption{\label{fig:5} The level-density $D(E_{S})$ versus
$E_{S}-E_{S_{min}}$\ (in $\hbar \omega $)\ with $\omega =100Hz$.
The lines from the shortest to longest are for $%
N=10^{2},\;10^{3},\;10^{4}$, and\;$10^{5}$, respectively. }
\end{figure}


\begin{thebibliography}{99}
\bibitem{stam98}  D. M. Stamper-Kurn, M. R. Andrews, A. P. Chikkatur, S.
Inouye, H. J. Miesner, J. Stenger, and W. Ketterle, Phys. Rev. Lett. \textbf{%
80}, 2027 (1998)

\bibitem{cira98}  J. I. Cirac, M. Lewenstein, K. Molmer, and P. Zoller,
Phys. Rev. A \textbf{57}, 1208 (1998)

\bibitem{duan02}  L. M. Duan, J. I. Cirac, and P. Zoller, Phys. Rev. A
\textbf{65}, 033619 (2002)

\bibitem{law98}  C. K. Law, H. Pu, and N. P. Bigelow, Phys. Rev. Lett.
\textbf{81}, 5257 (1998)

\bibitem{gold98}  E. Goldstein and P. Meystre, Phys. Rev. A \textbf{59},
3896 (1998)

\bibitem{ho00}  T. L. Ho and S. K. Yip, Phys. Rev. Lett. \textbf{84}, 4031
(2000)

\bibitem{ho98}  T. L. Ho, Phys. Rev. Lett. \textbf{81}, 742 (1998)

\bibitem{ohmi98}  T. Ohmi and K. Machida, J. Phys. Soc. Japan, \textbf{67},
1822 (1998)

\bibitem{zhan98}  W. Zhang and D.F. Walls, Phys. Rev. A \textbf{57}, 1248
(1998)

\bibitem{kemp02}  E.G.M. van Kempen \ et al, \ Phys. Rev. Lett. \ 88, 093201
(2002)

\bibitem{crub99}  A. Crubelier, \ et al, Eur. Phys. J. D \textbf{6}, 211
(1999)

\bibitem{koas00}  M. Koashi, and M. Ueda, Phys. Rev. Lett. \textbf{84}, 1066
(2000)

\bibitem{yi02}  S. Yi, \H{O}. E. M\H{u}stecapho\u{g}lu, C. P. Sun, and L.
You, Phys. Rev. A \textbf{66}, 011601R (2002)

\bibitem{bao03}  C. G. Bao, T. Y. Shi, Phys. Rev. A \textbf{68}, 032509
(2003)

\bibitem{bao02}  C. G. Bao, T. Y. Shi, Phys. Rev. A \textbf{66}, 013613
(2002)
\end{thebibliography}
\end{document}